\documentclass{svjour2}                    % onecolumn
\smartqed  % flush right qed marks, e.g. at end of proof
\usepackage{latexsym}
\begin{document}
\title{On an intrinsic quantum theoretical structure inside Einstein's gravity field equations.}
\subtitle{}
\author{J.F. Geurdes % etc
}                     % Do not remove
\offprints{han.geurdes@gmail.com}          % Insert a name or remove this line
\institute{}
\date{Received: 07-09-2010 / Revised version: date}
\maketitle
\abstract{
As is well known, Einstein was dissatisfied with the foundation of quantum theory and sought to find a basis for it that would have satisfied his need for a causal explanation. In this paper this abandoned idea is investigated. It is found that it is mathematically not dead at all. More in particular: a quantum mechanical U(1) gauge invariant Dirac equation can be derived from Einstein's gravity field equations. We ask ourselves what it means for physics, the history of physics and for the actual discussion on foundations.
} %end of abstract
\section{Introduction}
\label{intro}
Historical studies in physics are beneficial to establish our knowledge of nature. The historian can show the path that was followed by particular research groups or individuals and how this path lead to the present-day commonly accepted results. In addition, a philosopher or historian is allowed to wonder why certain ideas were not actualized in the past while others were. In the description of scientific advance, extra-scientific aspects like, the psychological, sociological and even economical, religious and political situation of the time can be incorporated. This embeds the activity of physical research in the historical context of society and add to an explanation of the reception of ideas. 

If one is interested in lines of research, the historical point of view also enables the possibility to take false leads of research serious. If there are e.g. rivaling concepts at certain times it is interesting to find reasons for the victory of the one over the other. In this sense history of physics somewhat resembles other branches of history like military history. In the latter one may ask the question what may have caused an army to have won or lose a battle. E.g. Napoleon lost the battle of Waterloo because of 'battle fatigue' generals. However, history of science is perhaps different because research ideas are not by necessity the same as strategies and decisions of generals. Revisiting a research program could imply the possibility of finding a new structure within a theory. In this sense a historical description of science could enter the arena of the science under study itself. It also could point at the fact that e.g. victory was claimed too early by one party. Perhaps even that in a larger historical perspective there are no victories or defeats but that it is all a matter of reformulation of a point of view. 

The present study is such a study. It attempts to uncover the inner structure of the idea of locality and causality because the inner structure has, no doubt, determined its faith. If this is clarified then perhaps reasons can be found why the idea was abandoned or did not got a fair chance. The present paper only lays the groundwork for the historical investigation and merely formulates hypotheses for further investigation. 

\section{Classical and quantum}
\label{sec:2}
In the present research we start with a first step in the study of the inner structure of Einstein's idea for causality and locality in quantum mechanics. The question is, is causality and locality in quantum theory a possible reality or just a dream? Despite of the fact that most present-day physicists reject, with reason, the idea of extra hidden parameters in quantum mechanics, in a historical perspective, this question makes sense. It should be noted beforehand that asking a question like this is not equal to questioning quantum theory itself. The latter interpretation is unjustified because Einstein never doubted the use of  quantum theory. Einstein only doubted whether or not quantum theory is complete \cite{Pais} and \cite{Eins}. The present paper also is not about the possibility of extra hidden parameters in Bell's theorem \cite{Wheeler} or in Hardy's paradox \cite{Hardy}. The author has reported on extra hidden parameters elsewhere \cite{Ge}, \cite{Gea}, \cite{Geb} and the arguments will not be repeated here. The present paper is about the false lead in research to search for a classical physics interpretation for quantum mechanics. 

\subsection{In defense of dead research ideas}
\label{sec:2.1}
Before writing down formulas and demonstrate the impossible, it is perhaps necessary to make some remarks. The author's defense of a search like this lies in his point of departure. This embraces the idea that science does not advance to an ever increasingly better understanding of the nature of things. The author's position to scientific progress is that science reformulates an empirically based position to fit explanations that are acceptable in a larger historical context.  Scientists are no super humans that stand above the influences of their times. They respond to it and most likely implicitly incorporate the historical situation to fit what is acceptable and needed as an 'explanation' of phenomena. One may wonder if the explanation in later periods must, by necessity, remain the same. Not only because new facts in nature are found and, by the way, one may wonder what the reasons are for these new facts. Explanations may also change in status or explaining power because of changing demand on an explanation. As an vide infra example one may think of the idea that classical physics refers to macroscopic object whereas in later times discoveries are made to favor the idea that there exist macroscopic bodies that behave quantum-like. In addition a certain type of explanation may lose its compelling power simply because more people tend to {\it believe} otherwise. The compelling power of a scientific explanation may decline because less people tend to believe it (be it right or wrong). Science is also a cultural penomenon.

A metahistorical point of view on science could be the possibility that knowledge already is present in certain older theories but did not make sense at that time or was not associated to existing problems. This latter point especially is vital to the paper as will become clear below. It suffice here to mention the fact that the 19-th century electromagnetic field theory of Maxwell contains interesting theoretical developments in the direction of quantum theory. This especially is true if one considers the fact that Maxwell  was aware of quaternions\footnote{Related to forces acting between two small magnets in reference to Professor Tait's work on quaternions} \cite{Max} while a relativistic free quantum wave equation can be formulated in a quaternion form \cite{DeLeo} \cite{Krav} and it was demonstrated that the Dirac equation can be derived from the classical electromagnetic field equations \cite{Ge2}, \cite{Sal}. In short, the intriguing idea is: could J.C. Maxwell have derived Dirac's equation himself from his e.m. field equations and have predicted e.g. the positron that was discovered in the 20-th century after the acceptance of Dirac's theory? Moreover, are there other classical theories that contain quantum aspects?

It is acknowledged that in this way the processes aimed to be described in nature by physics are put at considerable distance from human grasp. However, this will enable more room for the notion that explanations are man-made and based on creativity and extra-scientific impulses. How much nature there is contained in a certain theoretical description remains an open question that can only be answered by experiment. Some vital parts of theory -e.g. complex wave functions- are not open to direct experimental observation but are necessary in the theory to arrive at experimentally verified predictions.
\subsection{Where to search for a clue to structure}
\label{sec:2.2}
The unobservable parts of the quantum theory were a starting point of Einstein's questioning of the theory. In the beginning of quantum theory, Madelung \cite{Mad} derived a similarity between the Schr{\"o}dinger equation and hydrodynamics. This reformulation contained only real elements but those elements were embedded in a ether-like fluid structure that was difficult to acknowledge in a quantum domain. What was this fluid made off? Later David Bohm reformulated the hydrodynamical theory into the workings of a hidden quantum potential. The quantum potential in relation to quantum interpretation is excellently explained by Wigner \cite{Wigner}. In a relatively more recent past, Vaz and Rodrigues \cite{Vaz} and the present author \cite{Ge1}, \cite{Ge2} have demonstrated that classical field theory and modern relativistic quantum mechanics are related. This might justify the following search path. If relativistic quantum theory lies enclosed in classical electromagnetic field theory (see e.g.  Sallhofer \cite{Sal} too) then perhaps gravity theory is also a good candidate. In the following pages, this idea will be inspected further. It is stated beforehand that a no cosmological constant is incorporated in the theory. The presented result puts the search for a principal quantum gravity theory in a different perspective too.
\section{Equations}
\label{sec:3}
In the previous section (\ref{sec:2}) the context for the search was given. Here we will be involved in the mathematics and physics. 
\subsection{Preliminaries}
\label{sec:3.1}
Before embarking, let us first define the employed Minkowski metric, $\eta_{\mu,\nu} = diag(-1,1,1,1)$, or, $\eta_{1,1}=\eta_{2,2}=\eta_{3,3}=-\eta_{0,0}=1$, and, $\eta_{\mu,\nu}=0$, when, $\mu\neq\nu$, with, $\mu, \nu = 0,1,2,3$. Secondly, raising or lowering an index is performed with contraction using the Minkowski metric. For example, suppose, $a_{\mu}$ is a tensor, then 'raising the index' is done with, $a^{\lambda}=\eta^{\lambda,\mu}a_{\mu}$ and 
$\left(\forall : { \mu, \nu} = 0,1,2,3  \right) \eta^{\mu, \nu}=\eta_{\mu, \nu}$ element-by-element. Thirdly the to be investigated derivation is based on 'weak distortion' of the Minkowski metric. This means that  
\begin{equation}\label{e1}
g_{\mu,\nu} (x) = \eta_{\mu.\nu} + \sqrt{\epsilon} \varphi_{\mu,\nu}(x)
\end{equation}
with $\varphi_{\mu,\nu}(x) \sim O(\sqrt{\epsilon})$, terms of $O(\epsilon^2)$ will be repressed and $x=(x_0,x_1,x_2,x_3)$ spacetime coordinates.   Note that $O(h)$ is used in the sense of Landau's ordering symbol.
\subsection{Metric assumptions}
\label{sec:3.2}
The to be developed analysis is fairly general. However, below a specific example is given that meets the following conditions to satisfy possible criticism to the feasibility of the derived formal structure.  Let us in the notation for convenience leave out the separating comma if there can be no mistake in reading the indices. In the  derivation we will need the following facts. Firstly, $g^{\mu \nu}g_{\mu \nu}=\eta^{\mu \nu}\eta_{\mu \nu}=4$. From equation (\ref{e1}) it then follows that $\eta^{\mu \nu}\varphi_{\mu \nu}(x)=0$ and from the epsilonics it follows that $\epsilon \varphi^{\mu \nu}(x) \varphi_{\mu \nu}(x)\sim O(\epsilon^2)$ and can be suppressed. Secondly, we aim to have $g=-det(g_{\mu \nu})=1$. Let us suppose, for example, that $\left(\varphi_{\mu \nu}\right)=diag(\varphi_0,\varphi_1,\varphi_2,\varphi_3)$, with, to be even more specific, $\varphi_0(x)=\varphi_1(x)=f(x)$ and $\varphi_2=-h(x)$, $\varphi_3=h(x)$ and $f$ and $h$ two real functions. Then, the determinant equals
\begin{equation}\label{e2}
-g=  (-1+\sqrt{\epsilon} \varphi_0)(1+\sqrt{\epsilon} \varphi_1)(1+\sqrt{\epsilon} \varphi_2)(1+\sqrt{\epsilon} \varphi_3)
\end{equation}
We have, $\epsilon \varphi_{\mu}\varphi_{\nu}=O(\epsilon^2)$. Because $\varphi_2+\varphi_3=0$ and $\varphi_0=\varphi_1=f(x)$ we have $g^{\mu \nu}g_{\mu \nu}=4$. Moreover, suppressing $O(\epsilon^2)$
\begin{equation}\label{e3}
-g=  (-1+\sqrt{\epsilon} (\varphi_0 - \varphi_1))(1+\sqrt{\epsilon}( \varphi_2+\varphi_3))
\end{equation}
it follows, $g=1$. The reason for  $g=1$ will become clear later on. This example shows that there is a genuine metric $g_{\mu \nu}(x)$ that differs from the Minkowski metric and has the necessary characteristics. An additional assumption will be that $\partial_{\lambda}\varphi^{\lambda}_{~\nu}=a_{\nu}$, with, $a_{\nu}$ absolute constant. Now it should be noted that for, $\left(\varphi_{\mu \nu}\right)=diag(\varphi_0,\varphi_1,\varphi_2,\varphi_3)$, we have $\partial_{\lambda}\eta^{\lambda \sigma}\varphi_{\sigma \nu}=a_{\nu}$ and hence, $-\partial_0 \varphi_0 = a_0$ and  $\partial_k \varphi_k = a_k$, with, $k=1,2,3$. This further restricts the example metric $g_{\mu \nu}$ to
\begin{equation}\label{e3.1}
f(x) = a_1 x_1 - a_0 x_0 + f_{23}(x_2, x_3)
\end{equation} 
and, $f_{23}(x_2, x_3)$ containing the remainder in of space-time dependence of $f(x)$. Further, 
\begin{equation}\label{e3.2}
h(x)=a_3x_3 - a_2 x_2 + h_{01}(x_0,x_1)
\end{equation}
and similarly $h_{01}(x_0,x_1)$ the remainder in of space-time dependence of $h(x)$. Note that e.g. $\varphi_2 (x)= - h(x)$ and that in the previous further specification of $h(x)$ it is ensured that: $\partial_2 \varphi_2 = a_2$.  
\subsection{Field equations}
\label{sec:3.3}
As is well known, Einstein's field equations, relating the Ricci tensor, $R_{\mu \nu}(x)$, the stress-energy $T_{\mu \nu}(x)$ and the metric tensor, $g_{\mu \nu}(x)$, can be written as
\begin{equation}\label{e4}
R_{\mu \nu} = 8 \pi G \left( T_{\mu \nu}- {1\over{2}}g_{\mu \nu} T\right) 
\end{equation}
Note, $T=T(x)=g^{\mu \nu}(x)T_{\mu \nu}(x)$ and e.g. we use the notation $_0T(x)= \eta^{\mu \nu}(x)T_{\mu \nu}(x)$. In equation (\ref{e4}) $G$ is the gravitation constant, while it is assumed that $c=\hbar=1$.
The field equations are rewritten slightly for the convenience of the analysis. If $\kappa \sqrt{\epsilon}=8\pi G$, then,
\begin{equation}\label{e5}
R_{\mu \nu}=\kappa \sqrt{\epsilon} \left( T_{\mu \nu} -  {1\over{2}}g_{\mu \nu} T\right) 
\end{equation}
The Ricci tensor in equation (\ref{e5}) can be decomposed into $R_{\mu \nu} = r_{\mu \nu} + s_{\mu \nu}$. In terms of the affine connections the components of the Ricci tensor can be rewritten. For completeness:
\begin{equation}\label{e6}
\Gamma^{\lambda}_{\mu \nu}= {1\over{2}} g^{\lambda \sigma} \left(\partial_{\nu} g_{\sigma \mu} +\partial_{\mu} g_{\sigma \nu} -\partial_{\sigma} g_{\mu \nu} \right)
\end{equation}
The constituents of the Ricci tensor can subsequently be written in terms of the affine connections. For $r_{\mu \nu}$
\begin{equation}\label{e7}
r_{\mu \nu} = -\partial_{\lambda}\Gamma^{\lambda}_{\mu \nu} + \Gamma^{\sigma}_{\mu \alpha}\Gamma^{\alpha}_{\nu \sigma} 
\end{equation}
and $s_{\mu \nu}$
\begin{equation}\label{e8}
s_{\mu \nu} = \partial_{\nu} u_{\mu} - \Gamma^{\sigma}_{\mu \nu} u_{\sigma}
\end{equation}
Here $u_{\mu}$ is defined as a contraction on the affine connection related to $g=-det(g_{\mu \nu})$.
\begin{equation}\label{e9}
u_{\mu} = \Gamma^{\lambda}_{\lambda \mu } = \partial_{\mu} ln\left( \sqrt{g} \right) 
\end{equation}
We suppose that $g=1$ and in a previous section (section \ref{sec:3.2}) we saw that this is a genuine possibility among our other assumptions. In case $g=1$, we have $u_{\mu}=0$ and hence, $s_{\mu \nu}=0$. The field equations can then be rewritten as
\begin{equation}\label{e10}
r_{\mu \nu}=\kappa \sqrt{\epsilon} \left( T_{\mu \nu} -  {1\over{2}}g_{\mu \nu} T\right) 
\end{equation}
\subsection{Connection with QM}
\label{sec:3.4}
The basic equations for a derivation of Dirac's relativistic quantum equation will be given below. Because $R_{\mu \nu}$ is, using $g=1$, replaced by $r_{\mu \nu}$ in the field equations, according to equation (\ref{e7}) we need to inspect forms like $\partial_{\lambda}\Gamma^{\lambda}_{\mu \nu}$. Remembering the general form of the metric tensor in equation (\ref{e1}) we obtain the following
\begin{equation}\label{e11}
\partial_{\lambda}\Gamma^{\lambda}_{\mu \nu}={\sqrt{\epsilon}\over{2}}\eta^{\lambda \sigma}\left(\partial^{2}_{\lambda \mu} \varphi_{\sigma \nu}+\partial^{2}_{\lambda \nu} \varphi_{\sigma \mu} -\partial^{2}_{\lambda \sigma} \varphi_{\mu \nu}\right) 
\end{equation}
Note that the term $\epsilon \partial_{\lambda}Q^{\lambda}_{\mu  \nu} $ contained in $\partial_{\lambda}\Gamma^{\lambda}_{\mu \nu}$ is $O(\epsilon^2)$ because
\begin{equation}\label{e12}
2Q^{\lambda}_{\mu \nu} = \varphi^{\lambda \sigma}\left( \partial_{\mu}\varphi_{\sigma \nu}+\partial_{\nu}\varphi_{\sigma \mu}-\partial_{\sigma}\varphi_{\mu \nu}\right)
\end{equation}
and contractions like $\varphi^{\lambda \sigma}\partial_{\mu}\varphi_{\sigma \nu}$ are $O(\epsilon)$ because $\partial_{\mu } \varphi_{\sigma \nu}\sim O(\sqrt{\epsilon})$. Because of the additional assumption $\partial_{\lambda}\varphi^{\lambda}_{~\nu}=a_{\nu}$ and $a_{\nu}$ absolute constant suppressing $O(\epsilon^2)$ 
\begin{equation}\label{e13}
\partial_{\lambda}\Gamma^{\lambda}_{\mu \nu}=-{\sqrt{\epsilon}\over{2}}\left(\nabla ^2 - \partial^2_0\right)\varphi_{\mu \nu}
\end{equation}
From equation (\ref{e6}) and the definition of the metric in equation (\ref{e1}) we see that $\Gamma^{\lambda}_{\mu \nu}$ is $O(\epsilon)$. Hence, the product term of affine connections in the expression for $r_{\mu \nu}$ in equation (\ref{e7}) is of order $O(\epsilon^2)$ and can be suppressed. This leads to 
\begin{equation}\label{e14}
r_{\mu \nu}= {\sqrt{\epsilon}\over{2}}\left(\nabla ^2 - \partial^2_0\right)\varphi_{\mu \nu}
\end{equation}
From the field equations it then follows that
\begin{equation}\label{e15}
\Box^2\varphi_{\mu \nu}= 2\kappa \left( T_{\mu \nu} -  {1\over{2}}g_{\mu \nu} T\right)=k_{\mu \nu} 
\end{equation}
with $\Box^2=\left(\nabla ^2 - \partial^2_0\right)$ the D'Alembertian. From the previous equation, with $b^{\nu}$, possibly complex absolute constants, let us derive a vector $\phi_{\mu}=b^{\nu}\varphi_{\mu \nu}$ and $k_{\mu}=b^{\nu}k_{\mu \nu}$. With those two vectors $\phi_{\mu}$ and  $k_{\mu}$ in the relation $\Box^2\phi_{\mu} = k_{\mu}$, the present author derived a relativistic quantum-mechanical Dirac equation,
\begin{equation}\label{e16}
\gamma^{\mu}\left(\partial_{\mu}-A_{\mu}(x)\right)\psi(x)=\gamma^{\mu}D_{\mu}\psi(x)=0
\end{equation}
(see: Geurdes \cite{Ge2})\footnote{See the appendix also}. The 4x4 matirces $\gamma^{\mu}$ obey a Clifford algebra, $A_{\mu}(x)$ related to U(1) gauge and $\psi(x)$ a complex four vector.
\section{Meaning \& discussion}
\subsection{Physics consequences}
It is first and foremost stated that deriving a relativistic quantum equation from the (weak) gravity field equations is physically remarkable. In the first place because relativistic quantum theory can be derived from classical gravity. Secondly and because of a previous established relation between Maxwell's classical electromagnetic field equations, we now have obtained a theoretical relation between classical gravity and electromagnetic fields. This obtained relation is apparently not similar to gravity lensing known from astronomy because of the weak gravity field we employ. 

A second interesting point related to the derivation (see also the appendix section) is that a U(1) gauge transformation can be written as: $\psi \rightarrow e^{iR(x)}\psi$ with $R(x)$ a general real function of spacetime. In terms of the metric, when the transformed system is denoted with a prime, this leads to $\varphi'_{\mu \nu}(x)=e^{iR(x)}\varphi_{\mu \nu}(x)$ leading to complex valued $g'_{\mu \nu}(x)$ which is unphysical. The U(1) gauge then has to be restricted. An interesting possibility is to have $\theta(x_{\mu})=1$ when $x_{\mu}\geq 0$ and $\theta(x_{\mu})=0$ when $x_{\mu}<0$. Now if the gauge transformation is based on the following function
\begin{equation}\label{e17}
\iota(x_{\mu})=\sum_{n=-\infty}^{n=\infty }n \theta(x_{\mu} - n) \theta(n+1 - x_{\mu})
\end{equation}
with the gauge function $R$ as
\begin{equation}\label{e18}
R(x)=\pi \sum_{\nu=0}^{3}\iota(x_{\mu})
\end{equation}  
the transformed metric $g'_{\mu \nu} $ will obtain the following forms. For $x_{\mu}, \mu=0,1,2,3 $ such that $R(x)/\pi$ is even: $g'_{\mu \nu}(x)=\eta_{\mu \nu}(x)+\sqrt{\epsilon}\varphi_{\mu \nu}(x)$, while for, $R(x)/\pi$ is odd: $g'_{\mu \nu}(x)=\eta_{\mu \nu}(x)-\sqrt{\epsilon}\varphi_{\mu \nu}(x)$. In Newtonian form we have, the potential $V(x)\sim -\varphi_{0 0}(x)/2$ with the acceleration
\begin{equation}\label{e19}
\left({d^2 \vec{x}(t)\over{dt^2}}\right) = - \nabla V(x)
\end{equation}
Hence for $x_{\mu}, \mu=0,1,2,3 $ such that $R(x)/\pi$ is even, the acceleration after gauge transformation is equal to  $\left({d^2 \vec{x}(t)\over{dt^2}}\right )\sim \nabla \varphi_{0 0}(x)/2$. For $x_{\mu}, \mu=0,1,2,3 $ such that $R(x)/\pi$ is odd, the acceleration after gauge transformation is equal to $\left({d^2 \vec{x}(t)\over{dt^2}}\right) \sim -\nabla \varphi_{0 0}(x)/2$. Hence, when $x=(x_0,x_1,x_2,x_3)=(x_0,\vec{x})$ results in $R(x)/\pi$ is odd, this gives a reversal of the acceleration caused by the gauge transformation $\exp(iR)$. Perhaps this gauge transformation is unphysical but further (experimental) research appears to be necessary to verify such a claim. Note that the question can arise why complex wave functions are allowed when at the same time this kind of Gauges are rejected as physical.

The reader may note that if the mathematical transformation is 1-1 then a gravity field can be transformed into an electromagnetic field and vice versa. This claim is interesting and in need of further mathematical investigation but outside the scope of a historical investigation however. 

Note also that in a kind of second quantization scheme gravity theory can contain real quantum effects. This is perhaps a step, different from already known ones \cite{Rov} in the direction of a correct principal quantum gravity theory.
\subsection{History of physics ideas}
In a historical treatment of a false research idea one may wonder why the above obtained inclusion of Dirac's relativistic quantum equation was not discovered. Especially after 1990 when Sallhofer did his work on this relation there was perhaps reason enough to delve more deeper into classical and quantum field theory. Concerning the work of the author he only knows of references  related to electromagnetism \cite{Armour} and \cite{Luis}. Hence, the more general mathematical aspects of the relation were overlooked.

Suppose we employ the same reasoning for Einstein as we did for Maxwell previously. Then, Einstein could have had derived the Dirac equation from his field equations and one may wonder what the effect would have been on the course of physics research. Perhaps it is for the best that this did not happen but there definitely is a relation that cannot be ignored in a debate on the foundation of quantum theory. The assumptions in the section about the metric (section \ref{sec:3.2}) show that a situation in which this can occur is physically possible.

In the paper a false lead, namely the attempt to find a classical interpretation for quantum theory, leads to the surprising conclusion that gravity foield equations contain Dirac's relativistic quantum mechanical equation. In this sense it pays to, every now and then, revisit older ideas and to contrast them with already accepted ones. This is the first point the paper would like to make. The second point, supported by the electromagnetic field case, is that knowledge lies hidden in older theories. The hidden concepts in the 'classical' theories make little sense at the time but can later on be put in perspective. The author believes that this finding adds to the idea that science does not advance by producing increasingly better theories but by reformulating older views.

\section{Appendix: derivation of Dirac's quantum equation}\label{App}
In the sections below the derivation will be outlined. If the reader wants more detail he must consult \cite{Ge2}. In order to align with the previous analysis, we introduce $x_4=it$ and change $\mu$ etc running from 1 to 4 while having $g_{\mu,\nu}(\vec{x},x_4)$ for $\mu , \nu \in \{1,2,3,4\}$. This does not affect the conclusions but simply makes a comparison with \cite{Ge2} more easy.
\subsection{Definitions}\label{Appsub1}
Here the derivation of the relativistic quantum mechanical Dirac equation will be presented.  Let us write the complex vector equation (equation (\ref{e15}) and text):
\begin{equation}\label{a1}
\Box^2 \phi(x) = k(x)
\end{equation}
In the subsequent derivation use will be made of the matrices $M_1$ and $M_2$ that are defined with Kronecker delta's by $(M_1)_{\mu \nu}=\delta_{\mu \nu}\left(\delta_{\mu 1}+\delta_{\mu 2}\right)$ and $(M_2)_{\mu \nu} =\delta_{\mu \nu}\left(\delta_{\mu 3}+\delta_{\mu 4}\right)$. In the use of $M_a, (a=1,2)$ we will write $u_{b a} = M_a u_b$.   Now if we note that $\slash \hspace{-.25cm}D = \gamma^{\mu}D_{\mu}$ and $\gamma^{\mu}$ constituing a Clifford algebra and $\otimes$ the direct product of two four-vectors, $(\psi_1 \otimes \psi_2)^{\mu}=\psi^{\mu}_1\psi^{\mu}_2$, we may write for $\phi$
\begin{equation}\label{a2}
\phi(x)=\sum_{a,b,c,d=1}^2 \left(1-\delta_{b d}\right) \delta_{a c}\left(u_{\delta_{2 b}+2\delta_{1 b},a}- {v_{b,a}\over{2}} \right)\otimes v_{d c}
\end{equation}
For $k(x)$ we have
\begin{equation}\label{a3}
k(x)=-\slash \hspace{-.25cm}D \sum_{\mu = 1}^4 \gamma^{\mu}D_{\mu}h^{\mu} -(\slash \hspace{-.25cm}A)^2\phi + \gamma^{\mu}\gamma^{\nu}\left(\partial_{\mu}A_{\nu}  + \partial_{\nu}A_{\mu}\right)\phi 
\end{equation}
In the previous, $u_a$, $u_{b a}$, $v_{b a}$ and $h^{\mu} $ are indexed four-vectors that depend on space-time and $D_{\mu}$ as defined previously. The vector $h^{\mu}(x)$ is defined by
\begin{equation}\label{a4}
h^{\mu}=\sum_{a,b,c,d=1}^2 \left(1-\delta_{b d}\right) \delta_{a c} \left(u_{\delta_{2 b}+2\delta_{1 b},a}- v_{b,a} - { w^{\mu}_{b a} \over{2} }\right) \otimes w^{\mu}_{d c}
\end{equation}
For completeness: $w^{\mu}_{b a}$ is a four-vector with components $(w^{\mu 4}_{b a},w^{\mu 1}_{b a},w^{\mu 2}_{b a},w^{\mu 3}_{b a})$. 

Suppose that the four-vectors $u_a$, $v_{b a}$ and $w^{\mu}_{b a}$ are related by
\begin{equation}\label{a5}
\mathcal{R}\left(\gamma^{\mu} M_a\right)u_{\delta_{2 b}+2\delta_{1 b}} = \mathcal{R}\left(\gamma^{f(\mu)}S_a\right)u_b + \mathcal{R}\left(\gamma^{\mu}\right)\left(v_{b a} + w^{\mu}_{b a}\right) 
\end{equation}
Here we have
\begin{equation}\label{a6}
Y=\mathcal{R}(X) \Leftrightarrow Y_{\mu \nu} = \sqrt{X_{\mu \nu}}
\end{equation}
Moreover, the function $f$ is defined by $f(1)=2, f(2)=3, f(3)=1, f(4)=4$ and the matrix $S$ is defined by $S=I-\gamma^1 \gamma^2 - \gamma^2 \gamma^3 - \gamma^3 \gamma^1 = S_1 +S_2$  with $I$ the 4x4 unity matrix and $S\gamma^{\mu}=\gamma^{f(\mu)}S$. We have $(S_1)_{\mu \nu} = 0, (\mu \neq \nu)$. Furthermore, $(S_1)_{11} = (S_1)_{33}=1-i$ and $(S_1)_{22} = (S_1)_{44}=1+i $. For $S_2$ it is: $(S_2)_{12}=(S_2)_{34}=-1-i$ and $(S_2)_{21}=(S_2)_{43}=1-i$ while $(S_2)_{\mu \nu} =0$ for all other indices for $S_2$.

\subsection{Transformation of field equations}\label{AppTrans}
In the first place it can be easily verified that
\begin{equation}\label{a7}
\left[\mathcal{R}\left(\gamma^{\mu}M_a\right)u_1\right]\otimes\left[\mathcal{R}\left(\gamma^{\mu}M_a\right)u_2\right]=\gamma^{\mu} M_a \left(u_1 \otimes u_2\right)
\end{equation}
Substitution of equation (\ref{a5}) into (\ref{a7}) produces (for details see \cite{Ge2})
\begin{equation}\label{a8.0}
\begin{array}{lcl}
{\gamma^{\mu}M_a \left(u_1\otimes u_2\right)= \gamma^{f(\mu)}S_a \left(u_1\otimes u_2\right)}\\ 
{+ {1\over 2}\sum_{c,d=1}^2 \left( 1-\delta_{cd}\right)\gamma^{\mu} \left(v_{ca} + w^{\mu}_{ca} \right) \otimes \left(v_{da} + w^{\mu}_{da} \right) }\\
{+ \sum_{c,d=1}^2 }\left( 1-\delta_{cd}\right) \left[ \mathcal{R}\left(\gamma^{f(\mu)}S_a\right)u_c\right]\otimes \left[ \mathcal{R}\left(\gamma^{\mu}\right)\left( v_{da}+w^{\mu}_{da}\right)  \right]
\end{array}
\end{equation}
The term containing $\mathcal{R}\left(\gamma^{f(\mu)}S_a\right)u_c$ can be rewritten by using equation (\ref{a5}) again for $u_c$. Now if it is noted that $\mathcal{R}\left(\gamma^{\mu}M_a\right)u_c$ allows to employ $M_a$ outside the $\mathcal{R}$ operation giving $u_{ca}$ and introducing for brevity  $c^*=\delta_{2,c}+2\delta_{1,c}$ we then see from (\ref{a8.0})
\begin{equation}\label{a8.1}
\begin{array}{lcl}
{\gamma^{\mu}M_a \left(u_1\otimes u_2\right)= \gamma^{f(\mu)}S_a \left(u_1\otimes u_2\right)}\\ 
{-{1\over 2}\sum_{c,d=1}^2 \left( 1-\delta_{cd}\right)\gamma^{\mu} \left(v_{ca} + w^{\mu}_{ca} \right) \otimes \left(v_{da} + w^{\mu}_{da} \right) }\\
{ + \sum_{c,d=1}^2 \left( 1- \delta_{cd} \right)\gamma^{\mu} u_{c^*a} \otimes \left(v_{da} + w^{\mu}_{da} \right)   }
\end{array}
\end{equation}
Because, $I=M_1+M_2$ and $T=I-S$ and $S=S_1+S_2$, from equation (\ref{a5}) the following expression results
\begin{equation}\label{a8}
\begin{array}{lcl}
{ T \gamma^{\mu} \left(u_1 \otimes u_2\right) = \gamma^{\mu} \sum_{a,b,c,d=1}^2  (1-\delta_{cd})\delta_{ab} }\\
{\times \left[\left(u_{c^* a} - { v_{ca}\over{2} } \right)\otimes v_{db} + \left(u_{c^*a}-v_{ca} -{w^{\mu}_{ca}\over{2}}\right) \otimes w^{\mu}_{db}\right] }
\end{array}
\end{equation}
From the previous equation we can obtain the following simple expression employing the definitions provided for in the definition section (section \ref{Appsub1}), $T\gamma^{\mu}\left(u_1\otimes u_2 \right)=\gamma^{\mu}(\phi + h^{\mu})$ (no summation over $\mu$). Subsequent employing $D_{\mu}$ to this resulting equation gives
\begin{equation}\label{a9}
T\slash \hspace{-.25cm}D \left( u_1 \otimes u_2 \right) = \slash \hspace{-.25cm}D \phi + \sum_{\mu = 1}^4 \gamma^{\mu}D_{\mu}h^{\mu}
\end{equation}
If we subsequently apply $\slash \hspace{-.25cm}D$ to the left and right hand side of equation (\ref{a9}) then we obtain
\begin{equation}\label{a10}
\begin{array}{lcl}
{\slash \hspace{-.25cm}D T\slash \hspace{-.25cm}D\left( u_1 \otimes u_2 \right)= \Box^2 \phi + (\slash \hspace{-.25cm}A)^2 - }\\
{ \gamma^{\mu}\gamma^{\nu}\left(\partial_{\mu}A_{\nu}+\partial_{\nu}A_{\mu}\right)\phi + \slash \hspace{-.25cm}D  \sum_{\mu = 1}^4 \gamma^{\mu}D_{\mu}h^{\mu} } 
\end{array}
\end{equation}
From this it follows that $\slash \hspace{-.25cm}D T \slash \hspace{-.25cm}D (u_1 \otimes u_2)=0$ and with e.g. $\psi(x)=T \slash \hspace{-.25cm}D (u_1 \otimes u_2)(x)$, we find indeed, $\slash \hspace{-.25cm}D \psi(x)=0$ which belongs to quantum mechanics and represents a mass less particle in a gauge field. 
\subsection{Meaning of derivation}
In the previous section (\ref{AppTrans}) the relativistic Dirac equation was derived. This implies that hidden in a classical gravity field a mass less quantum equation can be found. This fact could point to the graviton idea. Its meaning is discussed elsewhere in the paper. 

Concerning the possible introduction of a mass containing term we reconsider, $\slash \hspace{-.25cm}D T \slash \hspace{-.25cm}D (u_1 \otimes u_2)=0$ and select $\slash \hspace{-.25cm}D (u_1 \otimes u_2)= T (\psi + \slash \hspace{-.25cm}D \chi)$. If, subsequently, $\slash \hspace{-.25cm}D^2 \chi = im\psi$, with $m$ the non-zero mass term, then because $T^2 = -3I_{4\times 4}$, it follows that: $\slash \hspace{-.25cm}D \psi + \slash \hspace{-.25cm}D^2 \chi = \slash \hspace{-.25cm}D \psi + i m \psi =0 $, which leads us to, $i \slash \hspace{-.25cm}D \psi -m \psi =0$. 
% BibTeX users please use
% \bibliographystyle{}
% \bibliography{}

\end{document}